\documentstyle[preprint,aps]{revtex}
\begin{document}
\draft

\title{OFF-DIAGONAL LONG-RANGE ORDER, RESTRICTED \\
GAUGE TRANSFORMATIONS, AND AHARONOV-BOHM \\
EFFECT IN CONDUCTORS}
\author{Murray Peshkin\cite{mppeshkin}}
\address{Physics Division, Argonne National Laboratory, Argonne, IL 60439-4843}

\maketitle
\begin{abstract}
The Hamiltonian describing a conductor surrounding an external magnetic
field contains a nonvanishing vector potential in the volume accessible to the
electrons and nuclei of which the conductor is made.  That vector potential
cannot be removed by a gauge transformation.  Nevertheless, a macroscopic
normal conductor
can experience no Aharonov-Bohm effect.  That is proved by assuming only that a
normal conductor lacks off-diagonal long-range order (ODLRO).
Then by restricting the Hilbert space to density matrices which lack ODLRO,
it is possible to introduce a restricted gauge transformation that removes
the interaction of the conductor with the vector potential.
\end{abstract}
\bigskip

The question has sometimes been raised as to whether the Aharonov-Bohm (AB)
effect
\cite{aharonov59,peshkin89} can be shielded by a conductor that surrounds
the magnetic field, as in Fig.\ 1a.  The beam particle induces charges and
currents in the conductor.  Those charges and currents may have their own AB
effect as they encircle the magnetic flux, and that may compensate the AB
effect on the beam particle.

There are also the more usual image charge and induced current effects, which
have nothing
to do with any interaction between the conductor and the external magnetic
field.
Image charges and induced currents act back on the beam particle and affect its
motion.
Those effects are not considered here.  They are negligible in current
experiments
on the AB effect.  In addition, they are at least quadratic in the charge of
the
beam particle, whereas the AB effect moves interference fringes proportionally
to
the charge of the beam particle for small fields.  [The limiting case of
diffraction
by a flux line of vanishing width is exceptional because the zero-flux
diffraction
vanishes in that limit.]

Experimentally \cite{peshkin89}, we know that AB effect is observed at its
full expected strength  although the magnetic field is always surrounded by a
conductor.
However, the beam particle typically has a velocity above 10$^{10}$ cm/sec and
the size
of the scattering center is typically microns, so the frequencies to which the
conductor
would have to respond would be of order 10$^{14}$ hz, approaching plasmon
frequencies in
metals, and one may speculate that shielding effects which may exist at lower
frequencies would not have been seen in the experiments performed to date
because
the conductors could not react quickly enough to the fields created by the fast
beam particles.  Experiments with slower beam particles would perhaps have a
better
chance to exhibit shielding of the AB effect because there a close-coupling
approximation, wherein the charge and current distributions in the conductor
follow the beam particle adiabatically around the conductor, should apply.
If such a phenomenon should exist for slower beam particles, it might raise the
possibility of using AB effect to probe properties of a macroscopic shield in
some
way analogous to the very productive experiments now done with mesoscopic
circuits.

The answer appears to be no; there can be no such shielding effect by a
macroscopic
conductor for beam particles of any energy.  That answer was given by A.S.
Goldhaber
\cite{goldhaber92}, both for normal and for superconducting conductors.
For superconducting shields, the key point is the flux quantization.
In the presence of a superconducting shield, the magnetic flux must be a
multiple of
$hc/2e$, half of LondonÕs unit.  However, the charge carriers have effectively
charge $2e$.
Therefore the AB phase shift of the superconducting electrons,
$(2\pi )\times$(charge)$\times$(flux), equals $2\pi$ and gives rise to no
observable effect.

For normal shields, Goldhaber's analysis relies upon specific and rather
subtle dynamical properties of the conductor which may not be general.  Here I
give
a proof that relies only on the most general property of normal matter, that it
does
not exhibit off-diagonal long-range order (ODLRO)
\cite{yang62}.  The conduction electrons do not have a coherent phase around
the ring and therefore cannot exhibit any AB effect of their own.  In other
words, the effects of the external flux on the dynamics of the conductor can be
removed by a gauge transformation even though the vector potential cannot be
removed by a gauge transformation.  That statement has been made before
\cite{peshkin}
in a speculative way.  Here I shall prove it.

To be gauge invariant, the Hamiltonian for the entire system must have the form

\begin{equation}
H_{\bf A} = H \left( {\bf X,P} - \frac{q}{c} {\bf A}({\bf X}),{\bf S,x}_j ,
{\bf p}_j
- \frac{e_j }{c} {\bf A}({\bf x}_j ), {\bf s}_j \right)
\end{equation}

The vector potential $\bf A$, assumed to be curl-free everywhere inside the
conductor, is that due to the external magnetic field.  Mutual magnetic
interactions of the particles are to be expressed as functions of their
dynamical variables.  $\bf X$, $\bf P$, and $\bf S$ are the coordinate,
canonical momentum,
and spin of the beam particle.  The ${\bf x}_{j}$, ${\bf p}_{j}$,
and ${\bf s}_{j}$ are the coordinates,
canonical momenta, and spins of all particles in the shield, electrons and
nuclei.
For an electron, the charge $q$ or $e_j$ is negative.

The vector potential cannot be removed by a gauge transformation, except for
special
values of the magnetic flux $\Phi$, because it must obey

\begin{equation}
\oint {\bf A} \cdot d\,{\bf r} = \Phi \,.
\end{equation}
The exceptional cases are those for which the flux obeys

\begin{equation}
\Phi = n \,\frac{hc}{e}
\end{equation}
with integer $n$.

If the conductor is simply connected, as in Fig.\ 1b, the interaction between
the magnetic flux and the particles in the conductor can be removed from the
Hamiltonian by a gauge transformation $U$ in the standard way.  Within the
domain of the Hamiltonian, i.e.\ when the coordinates ${\bf x}_j$ lie within
the
split-ring conductor of Fig.\ 1b,

\begin{eqnarray}
\Psi ' \left( {\bf X},\xi , {\bf x}_j , \xi_{j} , t \right) &=& \bar{U} \,\Psi
\left( {\bf X}, \xi, {\bf x}_j , \xi _j , t \right)
\nonumber \\
\bar{U }&=& \prod _{j} \, U({\bf x}_j )
\nonumber \\
U({\bf x}_j ) &=& exp \left\{ \frac{ie_{j}}{\hbar c} \int^{{\bf x}_j} \,
{\bf A}({\bf r}) \cdot d\,{\bf r} \right\}\,,
\end{eqnarray}
where $\xi$ and $\xi_{j}$ are the values of $S_z$ and $s_{jz}$.

\begin{equation}
\bar{H}_{\bf A} = \bar{U} H_{\bf A} \bar{U}^{-1} = H \left( {\bf X,P} -
\frac{q}{c}
{\bf A}({\bf X}), {\bf S}, {\bf x}_j , {\bf p}_j , {\bf s}_j \right) \,.
\end{equation}
The interaction between the external field and the beam particle is retained in
Eq.\ (5) through ${\bf A}({\bf X})$.

The density operator $\rho$, which, along with $H$, determines the dynamics,
obeys

\begin{equation}
\bar{\rho} = \bar{U} \rho \,\bar{U}^{-1} \,.
\end{equation}
Equivalently, the density matrix obeys\footnote{Following Ref.\ \cite{yang62},
the particles are in effect numbered and the statistics are imposed
through the symmetry of the density matrix.  For instance, if particles 1 and 2
are both electrons, then $\rho$ changes sign under
$({\bf x} _1 , \xi _1 ) \Leftrightarrow ( {\bf x} _2 , \xi _2 )$ and the
same is true of the primed variables.}

\begin{eqnarray}
\langle {\bf X},\xi, {\bf x}_1 , \xi _1 , \,\cdots \, {\bf x}_N , \xi _N
|\bar{\rho} (t) |
{\bf X}' , \xi ' , {\bf x}'_1 \xi'_1 , \,\cdots \,{\bf x}'_N , \xi '_N \rangle
=  \,\,\,\,\,
\nonumber \\
\bar{V} \langle {\bf X} ,\xi ,{\bf x}_1 , \xi _1 , \,\cdots\, {\bf x}_N , \xi
_N
|\rho (t) | {\bf X}', \xi ' , {\bf x}' _1 \xi '_1 , \,\cdots \,{\bf x}' _N ,
\xi '_N \rangle
\end{eqnarray}

\begin{equation}
\bar{V} = \prod_{j} V({\bf x}_j ,{\bf x}'_j ) = \prod_{j} exp \left\{
\frac{ie_j }{\hbar c}
\int^{{\bf x}'_j }_{{\bf x}_j } {\bf A}({\bf r}) \cdot d\, {\bf r} \right\}\,.
\end{equation}

For a simply-connected conductor, Eqs.\ (5) and (6) suffice to show that the
action of
the external magnetic field on the particles in the conductor is removed by a
gauge transformation and therefore the external field has no physical effect.
For a multiply-connected conductor such as the one in Fig.\ 1a, that proof
fails because the unitary operator $U$  does not exist except for values of
the magnetic flux that obey Eq.\ (3).  For all other values of the flux, the
function
$U({\bf x}_j )$ is multiple valued and it cannot carry a wave function within
the domain
of $H$ into a second wave function within the domain of $H$.  Similarly,
$V({\bf x}_j , {\bf x}'_j )$ is
multiple-valued and cannot carry an acceptable density matrix into a second
acceptable density matrix.  The multiple valuedness can be removed by making a
mathematical cut, for instance at the azimuthal angle $\phi = 0$, so that the
line
integrals of ${\bf A}$ become single valued, but then the wave functions become
discontinuous and the domain problem does not go away.

However, for a macroscopic normal conductor, the proof can be rescued by
restricting the
space of the density matrices to those which do not have ODLRO.  Strictly,
such density matrices obey

\begin{equation}
\lim_{|{\bf x}_j -{\bf x}'_j | \rightarrow \,\infty}
\langle {\bf X},\xi ,{\bf x}_1 , \xi _1 , \,\cdots \,{\bf x}_N , \xi _N |\rho
|{\bf X}' ,
\xi ' , {\bf x}'_1 \xi' _1  , \,\cdots \,{\bf x}'_N , \xi '_N \rangle = 0
\end{equation}
for each $j$ individually.  I will take a macroscopic normal ring to be one for
which

\begin{eqnarray}
\langle {\bf X},\xi , {\bf x}_1 , \xi _1 , \,\cdots \, {\bf x}_N , \xi _N |\rho
|{\bf X}' ,
\xi' , {\bf x}' _1 \xi' _1 , \,\cdots \,{\bf x} _N , \xi '_N \rangle = 0
\nonumber \\
{\rm when} \,\,|{\bf x}_j - {\bf x}'_j | >a \,\,\,{\rm for\,\, any} \,\, j
\end{eqnarray}
where $a$ is some length less than half
the length of the shortest path through the conducting ring that encircles the
magnetic flux.

Now each $\int^{{\bf x}'_j}_{{\bf x}_j} {\bf A}({\bf r})\cdot d\,{\bf r}$ in
Eq.\ (4)
can be made single-valued by requiring the integration path to obey

\begin{equation}
|{\bf r} - {\bf x}_j | < a \,\,\,\,\,\, {\rm {and}} \,\,\,\,\,\, |{\bf r} -{\bf
x}'_j | < a
\end{equation}
for every pair $({\bf x}_j , {\bf x}'_j )$ which obeys $|{\bf x}_j - {\bf x}'
_j | < a$.
It is unnecessary to define $V$ for other pairs, because the density matrix
in Eq.\ (7) vanishes for all those pairs.  Equations (7) and (8) define a
single-valued
density matrix $\bar{\rho}$ which is gauge equivalent to $\rho$.  There is no
discontinuity problem because $\rho$ vanishes in the regions where $V$ has a
jump in phase.

The same trick can be played on the Hamiltonian $H$.  The gauge transformation

\begin{equation}
\bar{H} = \bar{U}H\bar{U}^{-1}
\end{equation}
does not exist in general because it creates a multiple-valued Hamiltonian
that has no meaning, but in the truncated space of density matrices that
do not have ODLRO, that does not matter.  The matrix elements of $\bar{H}$ can
be defined
by the restricted gauge transformation

\begin{eqnarray}
\langle {\bf X},\xi ,{\bf x}_1 , \xi _1 , \,\cdots \, {\bf x}_N , \xi _N
|\bar{H} | {\bf X}',
{\bf x}'_1 \xi '_1 , \,\cdots \, {\bf x}' _N , \xi '_N \rangle =
\,\,\,\,\,\,\,\,\,\,\,\,\,\,\,
\nonumber \\
\bar {V} \langle {\bf X},\xi , {\bf x}_1 , \xi _1 , \, \cdots \,{\bf x}_N , \xi
_N
|H| {\bf X}' , \xi ', {\bf x}' _1, \xi ' _1 , \,\cdots \, {\bf x}'_N , \xi '_N
\rangle
\end{eqnarray}
whenever all pairs $({\bf x}_j , {\bf x}'_j )$ obey $|{\bf x}_j - {\bf x}'_j |
< a$.
Other matrix elements of $\bar{H}$ can be taken to vanish because they only
multiply vanishing matrix elements of the density matrix.  The
multiple-valuedness
problem has been eliminated and once again the interaction of the external
magnetic
field with the particles in the conductor has been removed from the Hamiltonian
and the density matrix.

The assumption that the density matrix exhibits no off-diagonal long-range
order at any time implies the assumption that the Schroedinger equation

\begin{equation}
i\hbar \,\frac{\partial\rho}{\partial t} = \left[ H,\rho \right]
\end{equation}
preserves the absence of ODLRO.  This proof would therefore not apply to
the unlikely situation where the passage of the beam particle somehow jostles
the
conductor into a superconducting state.

For mesoscopic circuits, on the nanometers scale, this proof fails because the
dimensions of the circuits are smaller than the length $a$ which measures the
range
of the off-diagonal order.  Finding the circuit size beyond which measured AB
effects in the conductor disappears might give a direct, albeit only
semi-quantitative,
measure of $a$.

\section*{Acknowledgment}
I thank F. Coester for useful discussions.
This work is supported by the U.S. Department of Energy, Nuclear Physics
Division, under contract W-31-109-ENG-38.

\begin{figure}
\caption{A conductor (shaded) surrounding a magnetic field region (black).\,\,
a) Intact, multiply connected, ring.\,\, b) Split, simply connected, ring.}
\end{figure}

\end{document}